\documentclass[reprint,
superscriptaddress,
 amsmath,amssymb,
 aps,
pre,
]{revtex4-2}

\usepackage{graphicx}
\usepackage{dcolumn}
\usepackage{bm}
\usepackage{xcolor}

\begin{document}

\preprint{APS/123-QED}

\title{Quantum thermodynamics of periodically driven polaritonic systems}

\author{Maicol A. Ochoa}
\email{maicol@umd.edu}
\affiliation{Department of Chemistry and Biochemistry, University of Maryland, College Park,MD, USA}


%
%

\date{\today}

\begin{abstract}

  We investigate the energy distribution and quantum thermodynamics in periodically driven polaritonic systems in the stationary state at room temperature. Specifically, we consider an exciton strongly coupled to a harmonic oscillator and quantify the energy reorganization between these two systems and their interaction as a function of coupling strength, driving force, and detuning. After deriving the quantum master equation for the polariton density matrix with weak environment interactions, we obtain the dissipative time propagator and the long-time evolution of an equilibrium initial state. This approach provides direct access to the stationary state and overcomes the difficulties found in the numerical evolution of weakly damped quantum systems near resonance, also providing maps on the polariton lineshape. Then, we compute the thermodynamic performance during harmonic modulation and demonstrate that maximum efficiency occurs at resonance. We also provide an expression for the irreversible heat rate and numerically demonstrate that this agrees with the thermodynamic laws.

\end{abstract}

\maketitle



\section{Introduction}\label{sec:int}
Hybrid polaritonic states, resulting from the interaction between excitons and photons, manifests on systems such as molecules in cavities\cite{li2021molecular,dunkelberger2022vibration}, plexcitons\cite{fofang2011plexciton,mondal2022coupling,mondal2022strong}, optically-stimulated semiconductors\cite{kaitouni2006engineering,deng2010exciton,ochoa2020extracting,khan2021efficient,denning2022cavity}, and nanomechanical devices\cite{pigeau2015observation,li2016hybrid,munoz2018hybrid}.  Excitonic polaritons display distinct properties that frequently depart from those of their components, providing an ideal platform for the investigation of strong coupling emergence, hybrid-state formation, their quantum control and potential quantum technologies\cite{kurizki2015quantum}. For foundational and practical reasons, it is interesting to study polaritons as nanoscale devices and, in particular, address their stability and performance under modulation due to external fields and forces.

Quantum thermodynamics\cite{pekola2015towards,esposito2015quantum,vinjanampathy2016quantum,deffner2019quantum,kosloff2019quantum} aims to extend thermodynamic concepts, such as work\cite{talkner2007fluctuation}, heat\cite{esposito2015nature,pekola2015towards,whitney2018quantum}, entropy\cite{esposito2015quantum,kalaee2021positivity}, and efficiency\cite{esposito2015efficiency} to nonequilibrium systems with a few degrees of freedom, evolving in regimes where quantum fluctuations cannot be ignored. \textcolor{black}{Studies in this field have concentrated on dissipative fermionic \cite{esposito2015quantum,ludovico2014dynamical,bruch2016quantum,haughian2018quantum} and bosonic\cite{ochoa2018quantum} systems under slow\cite{aurell2015neumann, li2017production,alicki1979quantum} or periodic modulation, and different techniques to describe periodically driven nanoscale systems are continuously emerging based on reduced quantum master equations\cite{hotz2021coarse,tanimura2020numerically,wacker2022nonresonant} and Floquet theory\cite{hone2009statistical,restrepo2018quantum,deng2016dynamics}}. Of particular interests are nanoscale systems and quantum materials resulting from light-matter interactions\cite{ochoa2015pump,elouard2020thermodynamics,wacker2022nonresonant}. Frequently, the dynamic and thermodynamic representation of these systems appear to be contradictory, especially when the interaction term, responsible for appearance of hybrid states, is of the same order of magnitude than the system energy\cite{ludovico2014dynamical,ochoa2016energy},  such as in  the case of strongly coupled polaritons. Recent studies in the polariton dynamics \cite{bamba2012dissipation,sieberer2013dynamical,drezet2017quantizing,pistorius2020quantum} attempt to account for dissipation in open or lossy systems. 

In this paper, we report on the quantum thermodynamics of periodically-driven dissipative polaritons, characterizing the stationary energy and population distribution. Our model describes weakly damped polaritons coupled to heat and work reservoirs, providing direct access to the stationary state and the absorption lineshape in terms of damping parameters, exciton-phonon coupling and detuning from the driving field frequency. Remarkably, our thermodynamic analysis reveals that the energy dissipated to the environment in the stationary state in the form of heat is related to the polariton von Neumann entropy.

\section{Polariton dynamics}\label{sec:dyn}
We consider a composite $\hat H_{\rm S}$ consisting of a two-level electronic system and a phonon, interacting with the environment $\hat H_{\rm B}$ and with interaction energy $\hat H_{\rm I} = \hat H_{\rm IX}+ \hat H_{\rm IP}$ given by ($\hbar = 1$)

\begin{align}
  \hat H =& \hat H_{\rm S} + \hat H_{\rm I} + \hat H_{\rm B} \label{eq:Htot}\\
  \hat H_{\rm S} =&   \sum_{i=1}^2 \varepsilon_i \hat d_i^\dagger \hat d_i + \omega \hat a^\dagger \hat a + V \hat d_2^\dagger \hat d_1 \hat a +V^* \hat a^\dagger \hat d_1^\dagger \hat d_2 \label{eq:Hsys}\\
  \hat H_{\rm IX} =& \sum_k W^X_{k} \hat c_k^\dagger \hat d_1 ^\dagger \hat d_2 + W^{X *}_{k} \, \hat d_2 ^\dagger \hat d_1 c_k \label{eq:HIX}\\
  \hat H_{\rm IP} =& \sum_k W^P_k \hat b_k^\dagger \hat a + \text{ h.c} \, \label{eq:HIP} ;
\end{align}
The Hamiltonian $\hat H$ in Eq.\ \eqref{eq:Htot} is reminiscent of the Jaynes-Cummings model and incorporates explicitly energy damping to the environment. In Eq.\ \eqref{eq:Hsys}, $\varepsilon_i$ is the $i$-th level energy, $\omega$ is the characteristic frequency for the phonon, $\hat d_i^\dagger $ ($\hat d_i$) is the creation (annihilation) operator for an electron in the $i$-th level, $\hat a^\dagger$ ($\hat a$) creates (annihilates) a phonon, and $V$ is the coupling term between the phonon and the two-level system. The electronic and phononic components of the system interact with the corresponding bosonic baths, namely,  $\hat H_X = \sum_k \varepsilon_k \hat c_k^\dagger \hat c_k$ and  $\hat H_P =\sum_k \omega_k \hat b_k^\dagger \hat b_k$, with coupling terms $W^X_{k}$ and $W^P_{k}$ as in Eqs.\ \eqref{eq:HIX} and \eqref{eq:HIP}. \textcolor{black}{These represent the heat reservoirs, which account for dissipative interactions with the environment. In the exciton case, these occur via radiative decay, energy damping via vibrational relaxation, or other nonradiative mechanisms. In the phonon case, dissipation occurs via couplings to different modes in the system or, in the case of a Fabry-Perot cavity, energy losses due to mirror leakage or absorption.}

As a consequence of the electron-phonon coupling, hybrid states $| \alpha \rangle$ are formed. These hybrid states diagonalize $\hat H_{\rm S}$, with corresponding eigenenergies $\lambda_\alpha$ and, in general, differ from the product states $|i , m \rangle$ characterized by the electronic quantum number $i$ and phonon mode $m$. To bring the polariton in a nonequilibrium condition, we consider the following generic periodic time-dependent variation, with frequency $\omega'$ and interaction strength $A$
\begin{align}
  \hat H_{d}(t) =& 2 A \cos(\omega' t ) \sum_{m} | 1, m \rangle \langle m, 2 | + | 2, m \rangle \langle m, 1 |,  \label{eq:driving}
\end{align}
and obtain a dynamical description of the driven polariton by solving the Liouville-von Neumann equation for the full density matrix. \textcolor{black}{The Hamiltonian $\hat H_d(t)$ describes the exciton interaction with an unspecified reversible work reservoir.\footnote{\textcolor{black}{A more general driving protocol will also include direct coupling to the phonon, e.g., in the following form
    \begin{equation*}
       2 A' \cos(\omega' t ) \sum_{i\in \{1,2\}} \sum_{m} | i, m \rangle \langle m+1, i| + | i, m+1 \rangle \langle m, i |, 
    \end{equation*}
    with interaction strength $A'$. It is possible to account for the dynamic and thermodynamic contributions of such interaction within the present formalism as this results in an additional time-dependent operator similar to $\mathcal{L}_d$, entering linearly in Eq.\ \eqref{eq:rhot}.}}} \textcolor{black}{Physically, the work reservoir can be an incident electromagnetic field coupled to the exciton via dipole-field interactions}. We solve the Liouville equation in the interaction picture with interacting Hamiltonian $\hat H_d(t)+ \hat H_I$, invoking the Born-Markov approximation\cite{breuer2002theory} (see Appendix \ref{ap:RME} for details). In this form, our solution is valid to second order in $W^X, W^P$, and $A$; holds for arbitrary values in the coupling term $V$. The resulting Markovian equation for the reduced density matrix $\rho(t)$ is
\begin{align}
  \frac{d}{dt} \vec \rho(t) = -\left(\mathcal{D} + \mathcal{L}_X+\mathcal{L}_P +\mathcal{L}_d(t)\right) \vec \rho(t) \label{eq:rhot},
\end{align}
where we write the reduced density matrix as a vector $ \vec \rho(t)$ with elements $\rho_\alpha^\beta (t) = \langle \alpha | \rho(t) |\beta \rangle$. In Eq.\ \eqref{eq:rhot}, the operator $\mathcal{D}$ is diagonal with matrix element  $\mathcal{D}_{\alpha_2 \beta_2}^{\alpha_1 \beta_1}   = -i (\lambda_{\alpha_1} -\lambda_{\beta_1}) \delta_{\alpha_1}^{\alpha_2} \delta_{\beta_1}^{\beta_2}$ and corresponds to the free evolution of the polariton in the absence of any relaxation mechanism;  $\mathcal{L}_X$ and $\mathcal{L}_P$ are operators that result from the independent exciton and phonon relaxation and are correspondingly proportional to the damping rates $\Gamma_\square = 2 \pi \sum_k |W_k^\square|^2 \delta (\omega_k - \omega)$ with $\square = X, P$. The time-dependent operator $\mathcal{L}_d(t)$ introduces the effects of the driving field in the polariton state, in a form that is proportional to $A^2$ and that depends on the frequency of the incident field $\omega'$. Notably, $\mathcal{L}_d(t)$ is periodic with period $\tau = 2 \pi /\omega'$ (see Appendix \ref{ap:RME} for details).

Formally, in the representation chosen in Eq. \eqref{eq:rhot}, the reduced density matrix has an infinite (countable) number of matrix elements. Due to the relaxation induced by the couplings to the environment, the contributions to the dynamics from higher energy states can be disregarded and a finite matrix representation of the operators in Eq.\ \eqref{eq:rhot}, in terms of the hybrid states formed by the lowest $m_o$ phonon modes, results in a faithful approximation to the dynamics. Within this finite representation, the $\mathcal{L}$ operators are $(2m_o)^2$-square matrices, and the time-dependent reduced density matrix obtained from the initial equilibrum condition $\vec \rho(0)$ is
\begin{align}
  \vec \rho(t) = \exp\left[-(\mathcal{D} + \mathcal{L}_X+\mathcal{L}_P) t  - \int_0^t \mathcal{L}_d(t') dt' \right] \vec \rho(0).\label{eq:rhoExp}
\end{align}
Utilizing the $\mathcal{L}_d$ periodicity we find that for $t = N \tau + \Delta t$,  $\Delta t < \tau $ 
\begin{align}
  \int_{0}^{t} \mathcal{L}_d(t') dt' =& \pi A^2 \left(\frac{t}{2}-i \frac{e^{2 i \omega' \Delta t}}{4 \omega'}\right) \mathcal{N}_+ \notag \\
  & + \pi A^2 \left(\frac{t}{2}+i \frac{e^{-2 i \omega' \Delta t}}{4 \omega'}\right) \mathcal{N}_{-}, \label{eq:LdB}  
\end{align}
where $\mathcal{N_\pm}$ are time-independent transition matrices describing the allowed system state conversions due to the external modulation.  \textcolor{black}{ We note that the operator in Eq.\ \eqref{eq:LdB} is a time-local matrix and, as a result, the evolution operator in Eq.\ \eqref{eq:rhoExp} is well-defined without a time-ordering operator.}

Equations \eqref{eq:rhoExp} and \eqref{eq:LdB} provide a simple description of a weakly damped periodically-modulated polariton in the strong coupling regime, as it holds for arbitrary values for the electron-phonon coupling strength $V$. Moreover, Eq.\ \eqref{eq:rhoExp} constitutes a numerically stable model for the dynamics of the driven system, naturally incorporating the effects of the system-bath couplings (system-bath coherences) within the Markovian regime, and allowing direct investigation of the polariton long-time evolution. In contrast, the numerical propagation of Eq.\ \eqref{eq:rhot} demands the initial evolution of the free system to introduce system-bath correlations and may not be stable near the resonance condition ($\omega = \omega'$), as in this case Eq.\ \eqref{eq:rhot} becomes a stiff differential equation.  In this form, Eqs.\  \eqref{eq:rhoExp} and \eqref{eq:LdB} overcome the difficulties found in the study of lossy cavities\cite{davidsson2020simulating} near resonance\cite{du2021nonequilibrium}, in the long-time evolution regime, and for an arbitrary driving frequency $\omega'$. Significantly,  Eqs.\ \eqref{eq:rhoExp} and \eqref{eq:LdB} are not limited to slow driving rates and do not require time-coarse graining\cite{hotz2021coarse} to describe the system in the long-time evolution limit.

\section{Polariton thermodynamics}\label{sec:ther}

Starting from Eq.\ \eqref{eq:rhot} we define element-wise the time-dependent operators $L_X, L_P$, and $L_d$ corresponding to the different partial time-derivatives of the polariton density matrix  
\begin{align}
  L^{\;\;\beta_1}_{\square\, \alpha_1} = \sum_{\alpha_2 \beta_2}\mathcal{L}^{\;\;\; \alpha_1 \beta_1}_{\square\, \alpha_2 \beta_2} \rho^{\beta_2}_{\alpha_2}(t)  \hspace{0.5cm}(\square = X, P , d),
\end{align}
and identify two reversible heat rates due to the polariton-heat reservoir interaction
\begin{align}
  \dot Q_X(t) =& - {\rm Tr} \left[ L_X(t) \hat H_{\rm S} \right ] \label{eq:dotQX}\\
  \dot Q_P(t) =& - {\rm Tr} \left[ L_P(t) \hat H_{\rm S} \right ]. \label{eq:dotQP}
\end{align}
Similar considerations were used in, e.g., Ref. \cite{wacker2022nonresonant} to define heat rates in dissipative driven TLS. Since we chose an interaction representation that includes the driving field  $\hat H_d(t) $ in the interaction term, the power or work rate on the polariton due to this external modulation is given by
\begin{align}
\dot W(t) =& - {\rm Tr} \left [ L_d (t) \hat H_{\rm S} \right ]. \label{eq:dotW}
\end{align}

 The polariton's von Neumann entropy 
\begin{align}
   S(t) = & - {\rm Tr } \left[ \rho(t) \ln \rho(t) \right],\label{eq:Stot}
\end{align}
significantly differs from the sum of the exciton and phonon residual entropies in the strong-coupling regime. However, the total entropy change rate is the linear combination of three different rates resulting from interactions with the environment
\begin{align}
  \dot S(t) =& \dot S_X(t)+\dot S_P(t)+\dot S_d(t), \\
  \dot S_{\square} =&  {\rm Tr } \left[ L_\square(t) \ln \rho(t) \right] \hspace{0.5cm} (\square = X, P, d) \label{eq:dotS}.
\end{align}
\textcolor{black}{Ideally, to consider the information entropy in Eq.\ \eqref{eq:Stot}, or any other, as the polariton nonequilibrium entropy function, its connection with the dissipated energy in the form of heat should closely follow the Clausius theorem.} Thus, in terms of the reversible heat and entropy rates, we define the effective polariton temperature $\beta_{\rm eff}$ as the proportionality factor between the total reversible heat rate and the sum of the associated entropy rates
\begin{align}
 \beta_{\rm eff} \left( \dot Q_X(t)+\dot Q_P(t)\right) =& \dot S_X(t)+\dot S_P(t) .\label{eq:Beff}
\end{align}
In this form, $\beta_{\rm eff}$  has units of inverse energy and may fluctuate as the driving field. \textcolor{black}{We assess the significance of $\beta_{\rm eff}$ by analyzing energy conservation}.  The first law imposes that the internal energy variation for the universe, $\dot U_{\rm univ}$, must always vanish. Consequently, $\dot U_{\rm S}+\dot U_{X}+\dot U_{P}+ \dot U_{W} = 0$, where $U_W$ is the work source internal energy. Since the thermal reservoirs are assumed to be reversible, then $\dot U_X = - \dot Q_x$, $\dot U_P = - \dot Q_p$. We identify an additional energy rate, $\beta_{\rm eff}^{-1}\dot S_d$, between the polariton and the work source. Assuming that the work source volume is fixed, we conclude that  $\dot U_W = - \beta_{\rm eff}^{-1}\dot S_d$.

\textcolor{black}{Finally, relative to the polariton equilibrium density matrix $\rho_o = e^{-\beta_o \hat H_{\rm S}}/{\rm Tr}[e^{-\beta_o \hat H_{\rm S}}]$, where $\beta_o = 1/(k_B T_{\rm env})$ and $T_{\rm env}$ is the environment temperature, we define the irreversible heat rate by invoking the Spohn theorem\cite{spohn1978entropy,alicki1979quantum,li2017production}  
\begin{align}
  \dot Q_{\rm irrev}(t) =& \beta_o^{-1} Tr[{(\mathcal{L}_X(t)+\mathcal{L}_P(t)) (\ln \rho_o - \ln \rho(t)\,)]} \notag\\
  =& \beta_o^{-1}(\dot S_X(t) + \dot S_p(t)) - (\dot Q_x(t) + \dot Q_P(t)). \label{eq:Qirrev}
\end{align}
}



\begin{figure}[t]
  \centering
  \includegraphics[scale=0.3]{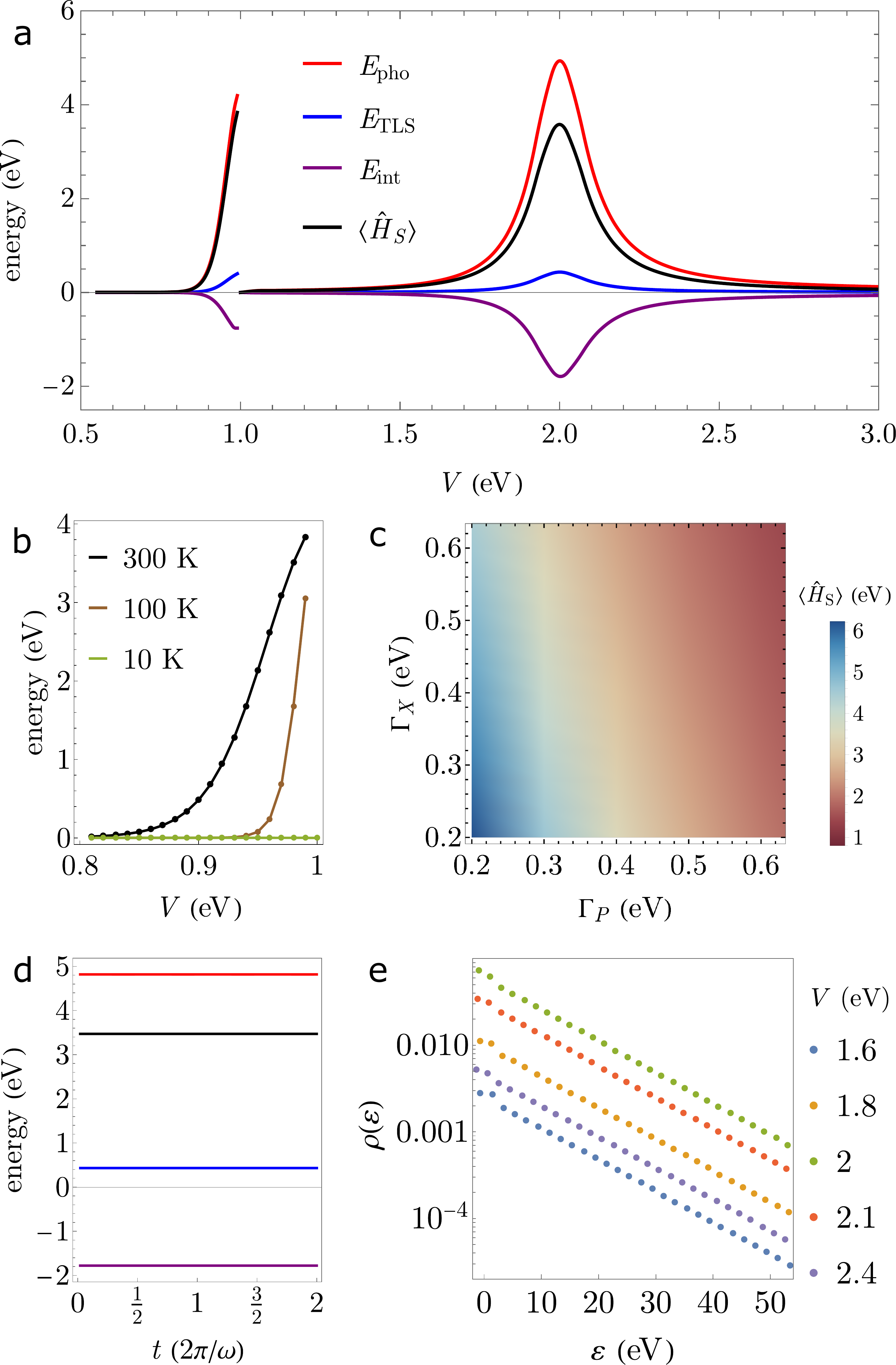}
  \caption{Polariton energy distribution. a) Stationary phonon, exciton, interaction, and polariton energies, respectively denoted by $E_{\rm pho}$, $E_{\rm TLS}$, $E_{\rm int}$, and  $\langle \hat H_{\rm S} \rangle$), as a function of coupling strength $V$. b) $\langle \hat H_{\rm S} \rangle$ near $V=\omega$ calculated at three different bath temperatures: 300 K (black), 100 K (brown), 10 K (green). c) $\langle \hat H_{\rm S} \rangle$ at resonance ($V= 2 \omega$) as function of the damping parameters $\Gamma_X$, $\Gamma_P$. d) Phonon, exciton, interaction and polariton energies as a function of time, and after a propagation period of 8.721 ps. (color code as in panel a ). e) Diagonal density matrix elements in the product state basis, as a function of energy and for several $V$.  Parameters are $\omega'=\omega =1.0$ eV, $\Gamma_X =0.2$ eV, $\Gamma_P =0.4$ eV, $A= 0.1$ eV, $\varepsilon_1= 0$ eV, $\varepsilon_2 = 1$ eV, $T=300$ K, unless otherwise specified.}\label{fig:lineshape}
\end{figure}


\section{Numerical example}\label{sec:num}
We now consider a driven dissipative polariton with identical driving field, phonon and exciton frequencies ( specifically $\omega' = \omega = \varepsilon_2 = 1$ eV, and $\varepsilon_1 = 0$ ) with driving parameter $A= 0.1$ eV, initially at the ground state (such that $\rho_\alpha^\beta(0) = \delta_{\alpha,(0,0)} \delta_{\beta,(0,0)}$ in the $|i,m \rangle$ basis). For this system, $\hat H_{\rm S}$ is a block matrix with eigenvalues $ \lambda_n^{\pm}= (n+1)\omega \pm V$ and zero. Figure \ref{fig:lineshape}a shows the phonon energy $E_{\rm pho} = \omega \langle \hat a^\dagger \hat a \rangle$, the exciton energy $E_{\rm TLS}= \varepsilon_2 \langle  \hat d_2^\dagger \hat d_2 \rangle$, the interaction energy $E_{\rm int} =  V \langle( \hat d_2^\dagger \hat d_1 \hat a +\hat a^\dagger \hat d_1^\dagger \hat d_2) \rangle $, and the total polariton energy $\langle \hat H_{\rm S} \rangle$ as a function of coupling strength $V$ at 300 K and after propagating the system 8.271 ps (2000$\tau$). In Fig.\ \ref{fig:lineshape} we assume the wideband limit and set the damping rates to $\Gamma_X =  0.2$ eV and $\Gamma_P =  0.4$ eV and include the lowest 30 phonon modes, which was enough to achieve numerical convergence in the rates(see Appendix \ref{ap:numconv} for additional details).  As we vary the coupling strength $V$, we find two significant features in the energy. First, we observe a monotonic increase in the absolute energies starting around $V = 0.85 $ eV with a sudden decay at 1 eV; and second, a broad peak near 2 eV. Indeed, when $V = \omega - \delta$, for small but positive $\delta$, the hybrid state $\lambda_0^- = \delta $ comes close in energy with the ground state, favoring the energy transfer cascade driven by thermal fluctuations. Figure \ref{fig:lineshape}b presents the polariton energy $\langle \hat H _{\rm S} \rangle$ near $V = \omega = 1 $ eV at 300, 100 and 10 K. We observe a reduction in the height and spread of this peak as the bath temperature drops, with a clear suppression of this peak at 10 K. This finding confirms that the energy absorption close to $V = \omega$ originates on the thermal fluctuations and the harmonic nature of the phonon. The second peak, centered at  $V = 2 \omega$, results from the emergence of a hybrid state with energy $\lambda_0^- = - \omega $, and the fact that the driving field frequency is in resonance with transition $\varepsilon_1 \to \lambda_0^{-}$. This resonant peak, can be fit to a Lorentzian distribution with a scale parameter linearly depending on $V$ and centered at $2$ eV (see Appendix \ref{ap:lineshape}). The energy contour in Fig.\ \ref{fig:lineshape}c, reveals how the total polariton energy in resonance varies for different values in the damping rates $\Gamma_X$ and $\Gamma_P$, suggesting that $\langle \hat H_{\rm S} \rangle$ is more sensitive to $\Gamma_P$ than $\Gamma_X$. We also note that majority of the polariton energy is stored in the phonon modes. Figure \ref{fig:lineshape}d shows that the system reaches a periodic stationary state. The energy oscillation amplitude in this stationary state is very small, on the order of a few $\mu$eV, as expected for an overdamped system with $A^2/{\rm eV} < \Gamma_X, \Gamma_P$. These oscillations are not detectable from Fig.\ \ref{fig:lineshape}d, and we can regard the polariton energy as constant. Figure \ref{fig:lineshape}e shows the density matrix diagonal elements for several values on the coupling strengths $V$, and as a function of the energy. These probabilities can be fit to a biexponential function with two energy decay parameters: one dominating at low energies, and the second one dominating at higher energies. Such biexponential fitting function is consistent with the definitions of $\hat H$ and $\hat H_d$ and the form for the density matrix in Eq.\ \eqref{eq:rhoExp}. Notably, while the population probabilities are maximal in resonance, their decay parameters do not change significantly with $V$.

\begin{figure}[t]
  \centering
  \includegraphics[scale=0.23]{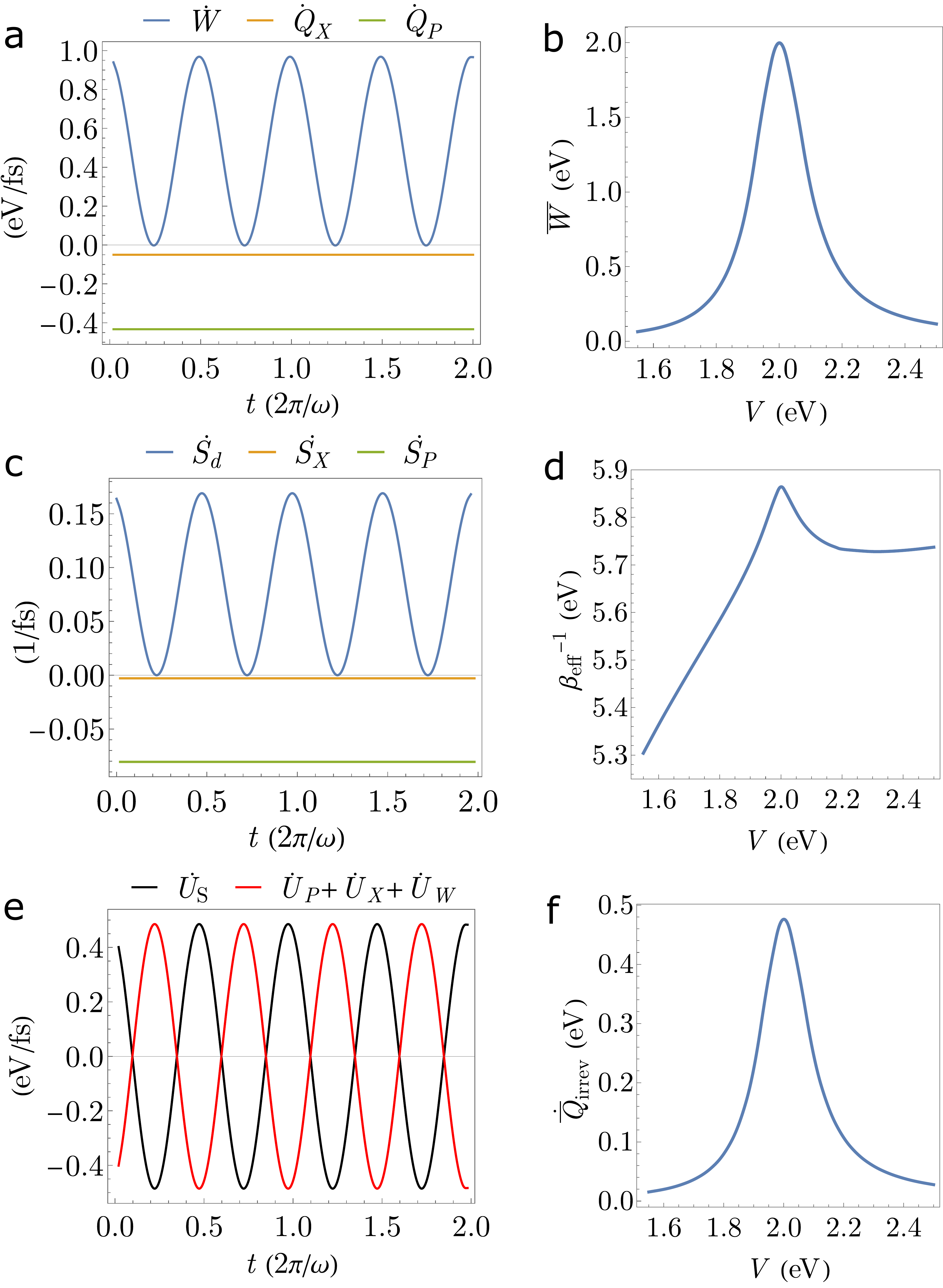}
  \caption{Polariton quantum thermodynamics. a) Work and heat rates as a function of time. b) Total work $\overline{W}$ done in the polariton by the external source per period $\tau$ as a function of coupling strength $V$. c) von Neumann entropy rates as a function of time. d) Inverse effective temperature $\beta_{\rm eff}^{-1}$ as a function of $V$. \textcolor{black}{e) Polariton $\dot U_{\rm S}$ and total reservoir $\dot U_P+\dot U_P+\dot U_P$ internal energy change rate as a function of time }f) Total irreversible dissipated heat $\overline{Q}_{\rm irrev}$ per period as a function of coupling strength $V$. These quantities are calculated after 8.271 ps of evolution. Parameters are as in Fig.\ \ref{fig:lineshape}.}\label{fig:rates}
\end{figure}

We investigate the polariton quantum thermodynamics under periodic driving in Fig.\ \ref{fig:rates}. First, we calculate the heat and work rates defined in Eqs.\ \eqref{eq:dotQX}, \eqref{eq:dotQP} and \eqref{eq:dotW} in the long-time evolution regime in Fig.\ \ref{fig:rates}a. We find negative and constant $\dot Q_X$ and $\dot Q_P$, indicating a net flux of heat energy from the polariton to the environment, while the work done on the system oscillates as the driving field. The total work per period $\overline{W} \equiv \int_{0}^{\tau} \dot W(t) dt$, presented in Fig.\ \ref{fig:rates}b serves as a measure of the thermodynamic efficiency of the process as a function of the coupling strength. Indeed, for a specific realization of the system, the intensity of the external field enters in the definition of the parameter $A$ in our model, which is kept constant in our calculations. The thermodynamic efficiency, defined by the ratio between $\overline{W}$ and the intensity of the external field, is therefore maximal in resonance ($\omega'=\omega$ and $V = 2 \omega$) as it is in these conditions that $\overline{W}$ is maximal. We note that during the cyclic process imposed by external modulation, the polariton remains in a stationary state far from equilibrium. As a result, the efficiency described here cannot be compared with the Carnot efficiency, and it is therefore the figure of merit for the performance of the external pumping. Figure \ref{fig:rates}c shows the von Neumann entropy rates defined in Eq.\ \eqref{eq:dotS} at resonance, and Fig.\ \ref{fig:rates}d presents the inverse effective temperature introduced in Eq.\ \eqref{eq:Beff} as a function of $V$. We observe that $\beta_{\rm eff}^{-1}$ achieves its maximum value at resonance, and it quickly decays away from this point.
\textcolor{black}{With this effective temperature, we numerically verify energy balance and the first law in the stationary state in Fig.\ \ref{fig:rates}e. We note that while $\dot U_S (t)$ oscillates on time, the environment's internal energy also oscillates, canceling the variation of the universe's internal energy at all times.} Finally, in Fig.\ \ref{fig:rates}f we show the total irreversible heat dissipated per period $\overline{Q}_{\rm irrev} \equiv \int_{0}^{\tau} \dot Q_{\rm irrev}(t) dt$ as a function of $V$, and find that this quantity is also maximal in resonance.

\section{Conclusion}
We developed a robust and efficient model for periodically driven polaritons, weakly interacting with external reservoirs, in the strong-coupling regime. We utilized this formalism to reveal the polariton long-time evolution and thermodynamic performance upon external modulation. Significantly, our approach is not limited to the stationary case and can also be used to investigate the polariton transient dynamics and performance. Moreover, we believe that our formalism permits the study of the quantum thermodynamics of other analog systems.\\

We thank the first reviewer for suggesting Ref.\ \citenum{alicki1979quantum} and the references therein.

\appendix

\section{Quantum Master Equation }\label{ap:RME}
We solve the Liouville-von Neumann (LvN) equation for the full density matrix $\bar \rho(t)$, assuming that the surroundings remain in equilibrium during the driving, such that,  $\bar \rho (t) = \rho^{\rm S} (t) \otimes \rho^{\rm B}_{\rm eq}$ at all times. We write the LvN equation in the interaction picture
\begin{align}\tag{A1}
  \frac{d \bar \rho_I}{d t} = -\frac{i}{\hbar}[\hat{\mathcal{V}}_I(t), \bar \rho_I (t)], \label{eq:qme1}
\end{align}
with interaction Hamiltonian
\begin{align}\tag{A2}
  \hat{\mathcal{V}}_I(t) = \hat H_I + \hat H_d(t),
\end{align}
and
\begin{align}\tag{A3}
  \bar \rho_I (t) =  e^{+i\hat H_{\rm S} t /\hbar}\, \bar \rho (t)  \, e^{-i\hat H_{\rm S} t /\hbar}
\end{align}

We solve Eq.\ \eqref{eq:qme1}, following the Born-Markov approximations described in, e.g., Ref.\ \citenum{breuer2002theory}.
First we integrate Eq.\ \eqref{eq:qme1}
\begin{align}\tag{A4}
  \rho_I(t) - \rho_I(0) = -\frac{i}{\hbar} \int_0^t [\hat{\mathcal{V}}_I(t'), \bar \rho_I (t')] \, dt' , \label{eq:qme2}
\end{align}
and replace Eq.\ \eqref{eq:qme2} in Eq.\ \eqref{eq:qme1} to obtain
\begin{align}
  \frac{d \bar \rho_I}{d t} =& -\frac{i}{\hbar}[\hat{\mathcal{V}}_I(t), \bar \rho_I (0)] \notag \\
                             &-\frac{1}{\hbar^2} \int_0^t [\hat{\mathcal{V}}_I(t), [\hat{\mathcal{V}}_I(t'), \bar \rho_I (t')]] dt' . \label{eq:qme3}\tag{A5}
\end{align}
Next, we trace out the bath degrees of freedom assuming independent baths and that the polariton is initially in an equilibrium state $\rho^S(0) = \rho_{\rm eq}$.
\begin{align}
\frac{d \rho_I^S}{d t} = &-\frac{1}{\hbar^2} \int_0^t {\rm Tr}_B\left\{[\hat{\mathcal{V}}_I(t), [\hat{\mathcal{V}}_I(t'), \bar \rho_I (t')]]\right\} dt' . \label{eq:qme4}\tag{A6}
\end{align}
and noting that
\begin{align}
  \frac{d \rho_I^S}{d t} = \frac{i}{\hbar} [\hat H_{\rm S}, \rho_I^S(t)] +   e^{+i\hat H_{\rm S} t /\hbar}\,  \frac{d \rho^S(t)}{d t}   \, e^{-i\hat H_{\rm S} t /\hbar}\tag{A7}
\end{align}
We finally obtain up to second order in $W_X$, $W_P$, and $A$  
\begin{align}
  \frac{d}{dt} \rho^{\rm S}(t) = &- \frac{i}{\hbar} [\hat H_{\rm S}, \rho^{\rm S}(t)] \notag\\
                                   &- \frac{1}{\hbar^2}\int_0^t e^{-i\hat H_{\rm S} t /\hbar}\times \notag \\
                                   &\hspace{2cm}{\rm Tr}_B \{ [\hat{\mathcal{V}}_I(t), [\hat{\mathcal{V}}_I(t'), \rho_I^{\rm S}(t')]]\} \notag \\
  & \hspace{5cm}\times e^{-i\hat H_{\rm S} t /\hbar}\, dt' \label{eq:LvNeq} \tag{A8}
\end{align}
From the first term in the right hand side in Eq.\ \eqref{eq:LvNeq} we recover $\mathcal{D}$.
\begin{align}
  \frac{\partial }{\partial t }\rho^{\alpha}_{\beta} (t) = -i (\lambda_{\alpha}-\lambda_{\beta}) \rho^{\alpha}_{\beta} (t). \tag{A9}
\end{align}
The second term incorporates the effect of the coupling of the polariton to the environment and the external driving, which are represented by the operators $\mathcal{L}_X$, $\mathcal{L}_P$, and $\mathcal{L}_d$ in the main text. Since the derivation of $\mathcal{L}_X$ and $\mathcal{L}_P$ follow a similar pattern, we illustrate this derivation in some detail for $\mathcal{L}_P$ and indicate how $\mathcal{L}_X$ results from simple substitutions.

\begin{figure}[t]
  \centering
  \includegraphics[scale=0.3]{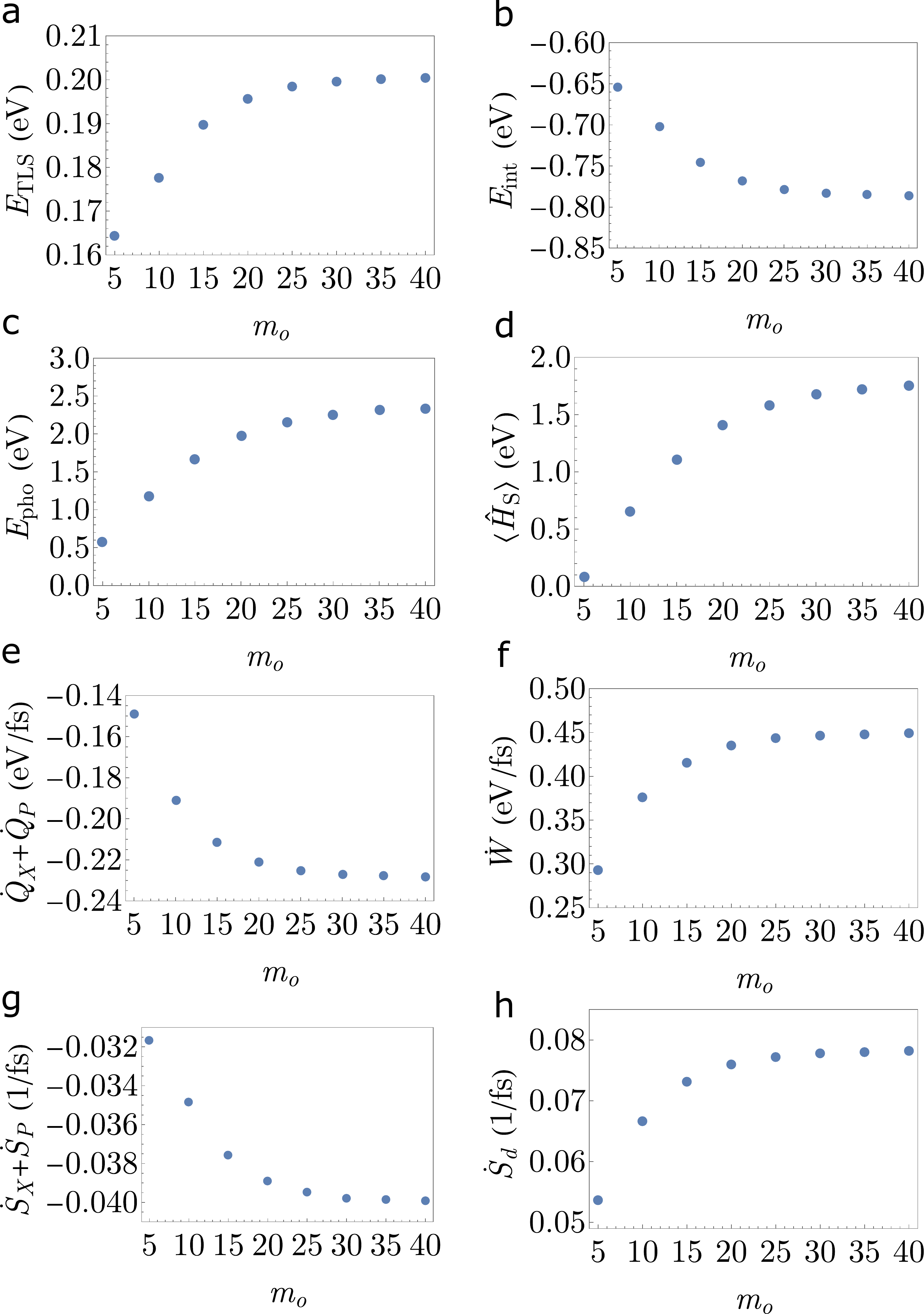}
  \caption{ Absolute energies, heat, work, and von Neumann entropy rates as a function of the upper bound in the number of phonon modes $m_o$, for the system in Figs.\ \ref{fig:lineshape} and \ref{fig:rates} , for $V= 1.9 \omega$, $\tau = 2000 \tau =8.271 $ ps, $\Gamma_P = 0.16$ e V $ = 4 \Gamma_X$. Other parameters are as in Figs. 1 and 2.}\label{fig:ConvOff}
\end{figure}

First, we introduce the self-energies 
\begin{align}
 \Sigma^{P,>} (t,t') =& \sum_k |W_k^P|^2 \langle \hat b_k(t)\hat b_k^\dagger(t')\rangle_B \tag{A10}\\ 
 \Sigma^{P,<} (t,t') =& \sum_k |W_k^P|^2 \langle \hat b_k^\dagger(t)\hat b_k(t')\rangle_B \tag{A11}
\end{align}
and adopt a Markovian approximation to the dynamics in Eq. \eqref{eq:LvNeq} by replacing $\rho_I(t') \to \rho_I(t)$. After this, we invoke the Redfield-Markov approximation and change the integral domain from $(0,t)$ to $(-\infty, \infty)$, with a rescaling $1/2$ factor such that  $\int_0^t \to \frac{1}{2}\int_{-\infty}^\infty $. The eight terms resulting from the explicit evaluation of the commutators in the second term in Eq. \eqref{eq:LvNeq} are
\begin{align}
  \mathcal{L}^{\;\;\; \alpha_1 \beta_1}_{P\, \alpha_2\beta_2}& =  \int_{-\infty}^\infty \tilde{\mathcal{L}}^{\;\;\; \alpha_1 \beta_1}_{P\, \alpha_2\beta_2}(t, t') dt' \notag \\
 =  \Gamma_P &\sum_\gamma \big( n(\gamma,\beta_1)\, \hat a^{\alpha_1}_{\gamma} \hat a^{\dagger \gamma}_{\;\; \alpha_2}  + (1 + n(\gamma,\beta_1))\, \hat a^{\dagger \alpha_1}_{\;\; \gamma} \hat a^{\gamma}_{\alpha_2}\big) \delta^{\beta_1}_{\beta_2}  \notag \\
  -\big( n&(\alpha_1,\gamma)\,\hat a^{\dagger \alpha_1} _{\;\; \alpha_2} \hat a^{\beta_2}_{\beta_1}  + (1 + n(\alpha_1,\gamma))\, \hat a^{\alpha_1}_{\alpha_2} \hat a^{ \dagger \beta_2}_{\;\; \beta_1} \big) \delta^{\beta_2}_{\gamma}  \notag \\
  +\big( n&(\gamma,\alpha_2)\,\hat a^{\beta_2} _{\gamma} \hat a^{\dagger \gamma}_{\;\; \beta_1}  + (1 + n(\gamma,\alpha_2))\, \hat a^{\dagger \beta_2}_{\;\; \gamma} \hat a^{\gamma}_{\beta_1} \big) \delta^{\alpha_2}_{\alpha_1}  \notag \\
   -\big( n&(\beta_1,\alpha_2)\,\hat a^{\dagger \alpha_1} _{\;\; \alpha_2} \hat a^{\gamma}_{\beta_1}  + (1 + n(\beta_1,\alpha_2))\, \hat a^{\alpha_1}_{\alpha_2} \hat a^{\dagger \gamma}_{\;\; \beta_1} \big) \delta^{\gamma}_{\beta_2} \tag{A12}
\end{align}
 with $\Gamma_P = 2 \pi \sum_k |W_k^P|^2 \delta(\omega_k - \omega)$. In this work we invoke the wideband approximation. This is justified under the assumption that $\Gamma_P$ is small compared to the width of the spectral function. As indicated above $\mathcal{L}_X$ follows from similar considerations, and under the substitution $\hat b_k \to \hat c_k$, $\hat a \to \hat d_1^\dagger \hat d_2$, and $W_P \to W_X$.

\begin{figure}[t]
  \centering
  \includegraphics[scale=0.3]{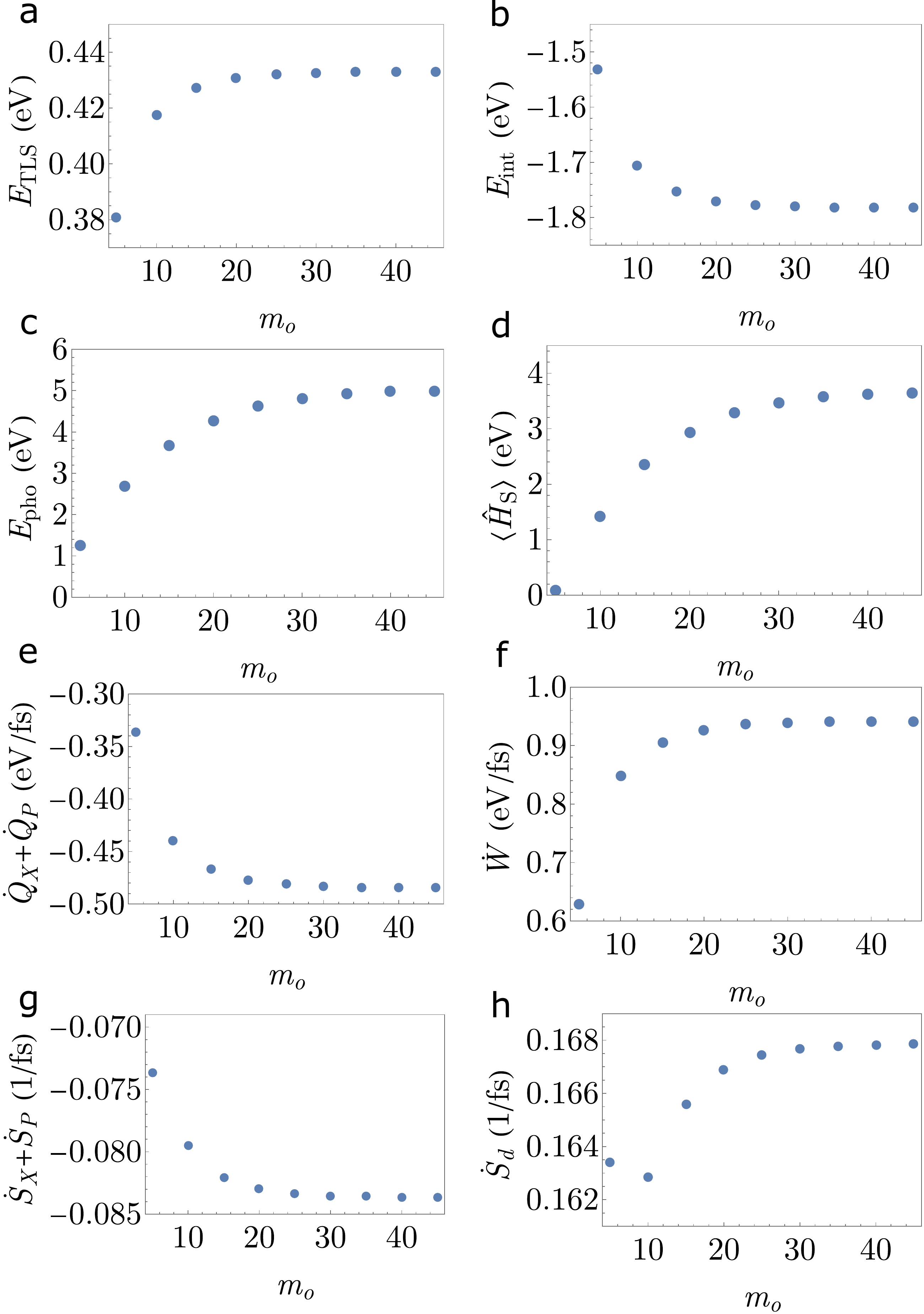}
  \caption{Absolute energies, heat, work, and von Neumann entropy rates as a function of the upper bound in the number of phonon modes $m_o$, for the system in Figs.\ \ref{fig:lineshape} and \ref{fig:rates}, for $V= 2 \omega$, $\tau = 2000 \tau =8.271 $ ps, $\Gamma_P = 0.16$ e V $ = 4 \Gamma_X$. Other parameters are as in Figs.\ \ref{fig:lineshape} and \ref{fig:rates}}\label{fig:ConvRes}
\end{figure}

 For the driving field introduced in Eq.\ \eqref{eq:driving} we write
 \begin{align}
   \hat x = & \hat d_1^\dagger \hat d_2  = \sum_m | 1, m \rangle \langle 2, m | \tag{A13}
 \end{align}
 and evaluate the double commutator in Eq.\ \eqref{eq:LvNeq}, under the Markovian approximations described above. For this we note that ($\hbar = 1$)
 \begin{align}
   \int_{-\infty}^{\infty}d t' &\cos(\omega t ) \cos(\omega t') e^{-i \Delta \lambda (t' -t)} = \notag\ \\
   & \pi \cos(\omega t) \left( e^{i \omega t} \delta(\Delta \lambda - \omega)+e^{-i \omega t} \delta(\Delta \lambda + \omega) \right), \tag{A14}
 \end{align}

 \noindent and as a result

 \begin{align}
   \mathcal{L}^{\;\;\; \alpha_1 \beta_1}_{d\, \alpha_2\beta_2}(t) = & -2 |A|^2  \cos(\omega' t) \times \notag\\
   \Big\{ & (g(\beta_1,\alpha_2,\omega', t)+g(\beta_2,\alpha_1,\omega',t))\notag\\
   &\hspace{1cm} \times \left(\hat x^{\dagger \alpha_2}_{\;\; \alpha_1} \hat x^{\beta_1}_{\beta_2} + \hat x^{\alpha_2}_{\alpha_1} \hat x^{\dagger \beta_1}_{\;\; \beta_2}  \right) \notag \\
   -\sum_\gamma & \Big( g(\beta_1,\gamma,\omega',t) \left(\hat x^{\dagger \gamma}_{\;\; \alpha_1} \hat x^{\alpha_2}_{\gamma} + \hat x^{\gamma}_{\alpha_1} \hat x^{\dagger \alpha_2}_{\;\; \gamma}  \right) \delta^{\beta_1}_{\beta_2} \notag\\
   & + g(\gamma,\alpha_1,\omega',t) \left(\hat x^{\dagger \gamma}_{\;\; \beta_2} \hat x^{\beta_1}_{\gamma} + \hat x^{\gamma}_{\beta_2} \hat x^{\dagger \beta_1}_{\;\; \gamma}  \right) \delta^{\alpha_1}_{\alpha_2}\Big) \Big\} \label{eq:Ldt} \tag{A15}
 \end{align}
 where
 \begin{align}
   g(\alpha, \beta, \omega, t) = &  e^{i \omega t } \delta(\lambda_{\alpha}-\lambda_\beta -\omega)+e^{-i \omega t } \delta(\lambda_{\alpha}-\lambda_\beta +\omega)\label{eq:gfunc}. \tag{A16}
 \end{align}
 From Eq. \eqref{eq:Ldt}, we obtain the transition matrices, $\mathcal{N}_{\pm}$, defined in Eq.\ \eqref{eq:LdB}
 \begin{align}
   \mathcal{N}&^{\;\;\; \alpha_1 \beta_1}_{\pm\, \alpha_2\beta_2} =   \notag\\ 
   &\Big\{  (  \delta(\lambda_{\beta_1}-\lambda_{\alpha_2} \mp \omega')+\delta(\lambda_{\beta_2}-\lambda_{\alpha_1} \mp \omega'))\notag\\
   &\hspace{2.5cm} \times \left(\hat x^{\dagger \alpha_2}_{\;\; \alpha_1} \hat x^{\beta_1}_{\beta_2} + \hat x^{\alpha_2}_{\alpha_1} \hat x^{\dagger \beta_1}_{\;\; \beta_2}  \right) \notag \\
   &-\sum_\gamma  \Big(  \delta(\lambda_{\beta_1}-\lambda_{\gamma} \mp \omega') \left(\hat x^{\dagger \gamma}_{\;\; \alpha_1} \hat x^{\alpha_2}_{\gamma} + \hat x^{\gamma}_{\alpha_1} \hat x^{\dagger \alpha_2}_{\;\; \gamma}  \right) \delta^{\beta_1}_{\beta_2} \notag\\
   &\hspace{0.3cm} + \delta(\lambda_{\gamma}-\lambda_{\alpha_1} \mp \omega') \left(\hat x^{\dagger \gamma}_{\;\; \beta_2} \hat x^{\beta_1}_{\gamma} + \hat x^{\gamma}_{\beta_2} \hat x^{\dagger \beta_1}_{\;\; \gamma}  \right) \delta^{\alpha_1}_{\alpha_2}\Big) \Big\}. \label{eq:Npm} \tag{A17}
 \end{align}

In our calculations involving Eqs.\ \eqref{eq:gfunc} and \eqref{eq:Npm}, we approximate the Dirac's delta function by a Lorentzian function with broadening parameter equal to $1 \times 10^{-4}$ eV.

\section{Numerical convergence tests}\label{ap:numconv}

In this section, we provide evidence for the numerical convergence of the energy and thermodynamic rates reported in Figs.\ \ref{fig:lineshape} and \ref{fig:rates}, when we carry out the summation over the phonon modes including only the first thirty modes ($m_o = 30$). In Figs. \ref{fig:ConvOff} and \ref{fig:ConvRes} we present the energies $E_{\rm TLS}$, $E_{\rm int}$, $E_{\rm pho}$, and $\langle \hat H_{\rm S} \rangle$; the heat and work rates $\dot Q_X+ \dot Q_P$ and $\dot W$; the von Neumann entropy rates  $\dot S_X+ \dot S_P$ and $\dot S_d$ correspondingly  off-resonance ($V = 1.9 \omega$) and in resonance ( $V = 2 \omega $) for $t = 2000 \tau =8.271$ ps. We find that the energy rates converge to the numerical exact value for $m_o= 30$. We also note that $E_{\rm pho}$ and $\langle \hat H_S \rangle$ deviate about 3 \% from the limit value in resonance ($m_o = 45$). This deviation is not relevant for the thermodynamic analysis.

\begin{figure}[h]
  \centering
  \includegraphics[scale=0.8]{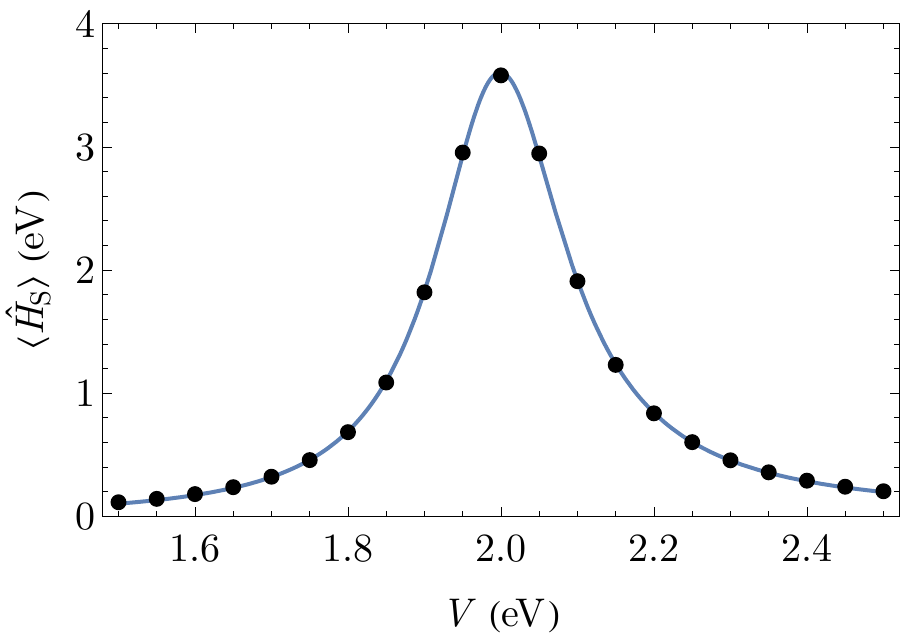}
  \caption{Polariton energy near $V = 2 \omega$, obtained from the density matrix (dots) and from the fitting function in Eq. \eqref{eq:fitline}. Other parameters are as in Fig.\ \ref{fig:lineshape}a}\label{fig:Lineshape}
\end{figure}

\section{Lineshape near Resonance}\label{ap:lineshape}

In this section we show that we can fit the second peak in Fig.\ 1a, centered at $V = 2 \omega$,  to a Lorentzian form with a broadening parameter that linearly depends on the exciton-phonon coupling strength. Explicitly, we write $\langle \hat H_S \rangle$ as function of $V$

\begin{align}
  \langle \hat H_{\rm S} \rangle = 0.376 * \frac{(0.0696 V - 0.0349)}{(V - 2.0)^2 + (0.0696 V - 0.0349)^2} \tag{C1} \label{eq:fitline}
\end{align}

Figure \ref{fig:Lineshape} presents a comparison between the energies obtained as the expectation value of $H_S$ and the ones obtained from the fitting function in Eq.\ \eqref{eq:fitline}.




\bibliography{PolaritonReferences.bib}

\end{document}